\documentclass[doublecol]{epl2} 
\usepackage{graphicx}
\usepackage{amssymb}
\usepackage{amsmath}
\usepackage{color}

\newcommand{\fig}[1]{Fig.~\ref{#1}}
\newcommand{\bs}[1]{\boldsymbol{#1}}                      % Vectors: bold
\newcommand\leftindex[2]{%                                % For left upper indexing, e.g. ~^2 A
  {\vphantom{#2}}#1#2%                                    %      --||--
}   

\title{Migration reversal of soft particles in vertical flows
}

\shorttitle{} %Insert here a short version of the title if it exceeds 70 characters

\author{Andre F\"ortsch \and Matthias Laumann \and Diego Kienle \and  Walter Zimmermann}
\shortauthor{A. F\"ortsch \etal}

\institute{Theoretische Physik I, Universit\"at Bayreuth, 95440 Bayreuth, Germany}

\abstract{
 Non-neutrally buoyant soft particles in vertical microflows are investigated.
We find,  soft particles lighter than the liquid
migrate to off-center streamlines 
in a downward Poiseuille flow (buoyancy-force antiparallel to flow). 
In contrast,  heavy soft particles 
migrate to the center of the downward (and vanishing) Poiseuille flow. 
A reversal of the flow direction causes in both cases a
{\it reversal of the migration direction}, i. e.   heavier (lighter) particles
migrate away  from (to) the center of a  parabolic flow profile.
Non-neutrally buoyant particles 
migrate also in a linear shear flow across the parallel streamlines:  heavy (light) particles migrate
along (antiparallel to) 
the local shear gradient. 
  This surprising, flow-dependent migration is characterized by simulations and analytical 
calculations for small particle deformations, confirming our  plausible explanation of the effect. This density 
dependent migration reversal may be useful for separating  particles. 
}

\pacs{47.15.G}{Low-Reynolds-number (creeping flows)}
\pacs{47.57.ef}{Sedimentation and migration}
\pacs{83.50.-v}{Deformation and flow}

\begin{document}

\maketitle

\section{Introduction}

Microfluidics is a rapidly evolving cross-disciplinary field, ranging 
from basic physics to a great variety of applications in life science and technology
\cite{Quake:2005.1,Kirby:2010,Nguyen:2010,Whitesides:2006.1,Popel:2005.1,Graham:2011.1,Beebe:2014.1,Kumar_S:2015.1,DiCarlo:2014.1}.
The blooming subfield of the dynamics  of {\it neutrally buoyant} soft particles in suspension and their cross-streamline
migration (CSM) in rectilinear shear flows, 
plays a central role for cell and DNA sorting, blood flow, polymer processing and so on 
\cite{Graham:2011.1,FrankeT:2014.1,Sajeesh:2014.1,Misbah:2014.1,Farutin:2016.1}. 
In contrast, little is known
about the dynamics of {\it non-neutrally buoyant} soft particles in rectilinear flows, but we  
show in this work for such particles a  novel migration reversal.

Segre and Silberberg reported in 1961 about  CSM of {\it neutrally buoyant} rigid particles at finite Reynolds numbers in flows 
through  pipes \cite{Silberberg:1961.Nat}.
When particles and channels approach the micrometer scale, fluid inertia does not matter 
and  particles follow the Stokesian dynamics. In this limit  CSM occurs only for soft particles but in
curvilinear~\cite{Shafer:1974.2,Nitsche:1996.1,Misbah:2011.1} 
as well as in
rectilinear flows ~\cite{Armstrong:1982.1,Brunn:1983.1,Freed:1985.1}, whereby 
in rectilinear flows, the flows fore-aft
symmetry is broken, requiring intra-particle hydrodynamic
interaction~\cite{Armstrong:1982.1,Brunn:1983.1}.
Such symmetry breaking occurs also near boundaries via wall-induced lift
forces \cite{Freed:1985.1,Misbah:1999.1,Seifert:1999.1,Viallat:2002.1,Graham:2005.1}
or by space-dependent shear rates, so that
dumbbells~\cite{Armstrong:1982.1,Brunn:1983.1},
droplets~\cite{Leal:1980.1,Chakraborty:2015.1},
vesicles and capsules~\cite{Kaoui:2008.1,Misbah:2008.1,Bagchi:2008.1}
exhibit CSM even in unbounded flow. Such parity breaking mechanisms may be also accompanied by 
 a viscosity contrast~\cite{Farutin:2014.1} or chirality ~\cite{LarsonRG:2009.1}. Recently 
 was found, that CSM takes place also for asymmetric soft particles in time-dependent linear shear flow \cite{Laumann:2017.1}
 and  that soft particles are actuated even in a homogeneous but time-dependent flow by taking particle inertia into account \cite{Kanso:2016.1}.

\begin{figure}[htb]
\vspace{-1mm}
\begin{center}
\includegraphics[width=0.98\columnwidth]{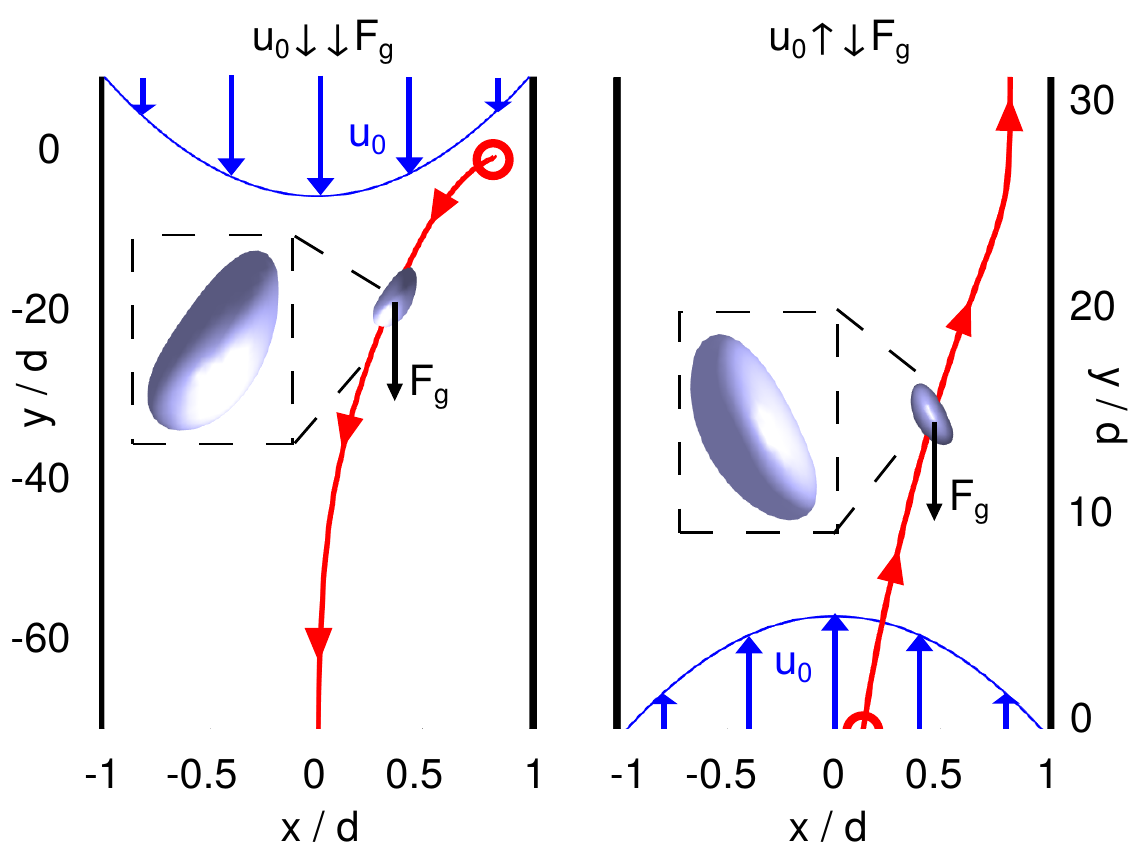}
\end{center}
\vspace{-4mm}
\caption{The left part shows in red the center of mass trajectory, $(x_c(t),y_c(t))$, of a heavy and soft tank-treadin capsule in a downward  Poiseuille flow,   $\bs u_0(x)$,  parallel to the
gravitational force $\bs F_g$: $ \bs u_0 \downarrow \downarrow \bs F_g$.
The  right part shows a  trajectory (red) of a heavy  capsule in an upward Poiseuille flow, i. e. 
$ \bs u_0 \uparrow \downarrow \bs F_g$, where the capsule migrates  away 
from the center of the Poiseuille flow.
The two confining walls are at a distance $2d$.
%
%
%in the left part and for an upward flow (right) part . 
%In the left part the gravitational force $\bs F_g$ is parallel to the 
%down flow  $\bs u_0(\bs r)$  
%($\downarrow \downarrow$) and the capsule migrates to the center of the Poiseuille flow. 
%In the right part gravitation is antiparallel to the up-flow,  $ \bs u_0 \uparrow \downarrow \bs F_g$,
%and the capsule migrates away from the center to a parameter dependent off-center position.
}
\label{Sketchsink}
\end{figure}

Heavy rigid particles  in a finite Reynolds number flow downward in a gravitational field migrate away from the tube center 
and for an upward flow to the tube center, as experimentally observed \cite{JeffreyRC:1965.1}.
Effects of axial forces on rigid particles  along the tube axis in finite Reynolds number flows where also
studied in Refs. \cite{KimJW:2009.1,Prohm:2015.1} and  effects of axial (electrical) forces on (charged) polymers in in pipe flows
in Refs. \cite{Yeung_ES:2002.1,Ladd:2007.1}. Little is known about CSM of non-neutrally buoyant soft particles in vertical Stokes flows.

Here we show that a soft heavy  particle migrates to the center of a tube in the limit of a vanishing Reynolds number, while  rigid particles don't \cite{Joseph:94.1}.
Furthermore is shown,  that   heavy (light)
soft particles migrate in vertical rectilinear Stokes flows antiparallel (parallel) to the shear gradient. 
This dependence of the CSM
direction on the shear gradient is shown by approximate analytical calculations and 
by numerical simulations  for soft capsules and
ring polymers. 
Also a plausible qualitative explanation of the  origin migration of non-neutral particles is provided: It is 
based on the  interplay between the 
orientation of the  shear induced elliptical shaped soft capsule (ring)  together with 
the related anisotropic friction, a  non-buoyant particle experiences.

%%%%%%%%%%%%%%%%%%%%%%%%%%%%%%%
\section{Modeling soft particles in Stokes flow}
\label{models}
%%%%%%%%%%%%%%%%%%%%%%%%%%%%%%%
The dynamical equations of  two non-neutrally buoyant particles, bead-spring models for  ring polymers and  elastic capsules
in rectilinear flows is described in this section.
The unperturbed linear shear flow is given by  $\bs u_0(x)=\dot \gamma x {\bf e}_y$ and the   Poiseuille flow  
between the two confining plane boundaries 
at $x_d=\pm d$ by
\begin{align}
\bs u_0(x) =\hat u_0 \left(1-\frac{x^2}{d^2}\right) {\bf e}_y\,.
 \label{eq_poiseuille_flow}
\end{align}
The maximal velocity
$\hat u_0$ at the center $y=0$ and the shear rate $\dot \gamma$ can be either positive or negative.

The migration  of the soft particles is obtained by their non-Brownian trajectories.
The trajectories of the bead-spring ring-polymer and the capsule in an unperturbed 
 $\bs u_0(x)$ are determined by solving the standard Stokesian dynamics for bead-spring models
with the position  $\bs r_i$ of the $i$-th bead:
\begin{align}
\label{beaddynamics}
	\bs{\dot r}_i = \bs u_0({\bf r}_i) + \sum \limits_{j=1}^N  \mbox{\bf{ H}}_{ij} \bs F_j\,.
\end{align}
$\bs F_j$ describes the force acting on the $j-$th bead and $\mbox{\bf{ H}}_{ij}$ is the mobility matrix described
in the following.

The  harmonic spring potential for a ring polymer with a finite mean distance $b$ between next-neighbor beads 
and spring constant $k$ is given by:
\begin{align}
	V_{spr} = \frac k 2 \sum \limits_{i=1}^N \left( b - r_{ij} \right)^2\,.
\end{align}
Due to the 
bending potential
\begin{align}
	V_b = -\frac \kappa 2 \sum \limits_{i=1}^N \ln\left( 1 + \cos \beta_i \right),
\end{align}
with the bending constant $\kappa$  the closed polymer has in the undeformed state 
the shape of a ring. The angle $\beta_i$ is given by 
$ \cos \left(\beta_i \right) = \hat{\bs{r}}_{(i-1)i} \cdot \hat{\bs{r}}_{i(i+1)}$ and the distance vector between the beads by $\bs{r}_{i,j}= \bs{r}_{i}-\bs{r}_{j}=r_{i,j}\hat{\bs r}_{i,j}$. Unit vectors are denoted by a hat.
The forces $\bs F_j$ in Eq.~(\ref{beaddynamics}) are given by $ \bs F_j = -\nabla_j \left[ V_{spr} + V_b\right]  + \bs F_g $,
with ${\bs F}_g=F_g {\bf e}_y$.

% Oseen Tensor 
$H_{ij}$  is the mobility matrix describing the hydrodynamic interactions between beads in 
the presence of a single wall parallel to  $yz$-plane with no-slip boundary condition,
which is of the following form \cite{Blake:1971.1}:
\begin{align}
	\label{eq:blake_tensor}
 H_{ij}(\bs r_i,\bs r_j)  &= \leftindex{^S}{H_{ij}} - \leftindex{^S}{H_{ij}}(\bs r_i, \bs r_j^\prime) \nonumber\\
 &+ \leftindex{^D}{H_{ij}}(\bs r_i,\bs r_j^\prime)-\leftindex{^{SD}}{H_{ij}}(\bs r_i,\bs r_j^\prime)\,.
\end{align}
Herein $\bs r^\prime_j=(x_j+2h_j,y_j,z_j)$ is the position of a  mirror-particle to the  $j$-th bead with distance $h_j$ to the wall. 
The first term contains the hydrodynamic interaction (HI) in the bulk regime, represented by the Oseen tensor \cite{Dhont:96}
\begin{align}
 \leftindex{^S}{H_{ij}^{\alpha \beta}}(\bs r_i,\bs r_j) = \left\{ \begin{array}{c c}
 \frac{1}{8\pi\eta r_{ij}}\left( \delta_{\alpha \beta} + \frac{ r_{ij}^\alpha r_{ij}^\beta}{r^2_{ij}} \right)  & i \neq j \\
 \frac{1}{6\pi\eta a} \delta_{\alpha\beta}
 \end{array}\right. ,
\end{align}
where $\alpha,\beta \in \{x,y,z\}$ and $6\pi\eta a$ is the Stokes friction with the 
bulk viscosity $\eta$ and bead radius $a$. The second term in Eq.~(\ref{eq:blake_tensor}) is the HI generated by the mirror image
\begin{align}
 \leftindex{^S}{H_{ij}^{\alpha \beta}}(\bs r_i,\bs r^\prime_j) = \frac{1}{8\pi\eta \tilde r_{ij}}\left( \delta_{\alpha \beta} + \frac{ \tilde r_{ij}^\alpha \tilde r_{ij}^\beta}{\tilde r^2_{ij}} \right) , 
\end{align}
with $\tilde{ \bs r}_{ij} = \bs r_i - \bs r_j^\prime = \tilde r_{ij} \hat{\tilde{\bs{r}}}_{ij}$ is the distance to the mirror bead image $j$. The last two terms in \eqref{eq:blake_tensor} contain the Stokes doublet (D) part
\begin{align}
	\leftindex{^D}{H_{ij}^{\alpha \beta}} (\bs r_i, \bs r_j^\prime) = \frac{h_j^2 ( 1 - 2\delta_{\beta y})}{4 \pi \eta \tilde r_{ij}^3}  \left( \delta_{\alpha \beta} - 3 \frac{\tilde r_{ij}^\alpha \tilde r^\beta_{ij}}{\tilde r^2_{ij}}\right),
\end{align}
and the source doublet (SD) part 
\begin{align}
	\leftindex{^{SD}}{H_{ij}^{\alpha\beta}}&(\bs r_i,\bs r_j^\prime) = \frac{1}{4\pi\eta\tilde r^3_{ij}} h_j \left( 1-2 \delta_{\beta y}\right) \nonumber \\
	&\left( \delta_{\alpha\beta} \tilde r^y_{ij} - \delta_{\alpha y}\tilde r^\beta_{ij} + \delta_{\beta y} \tilde r^\alpha_{ij} - 3\frac{\tilde r_{ij}^\alpha \tilde r_{ij}^\beta \tilde r_{ij}^y}{\tilde r^2_{ij}} \right) 
\end{align}
of the HI. To take account of the effects of  the second wall, a superposition of two single walls is used. This 
approximation generates according to \cite{Jones:2004.1} reasonable results, if the particle size to channel-width ratio is less than $5$. In simulations without walls only $\leftindex{^S}{\bs{ H}}_{ij}$ is used.

In case of the capsule we use the same equations of motion, but different potential forces.
The elastic forces of the capsule are described by the Neo-Hookean Law with Potential $V_{NH}$. 
The Neo-Hookean Law describes a rubber like material with a constant surface shear elastic 
modulus $G$ \cite{BarthesBiesel:1981.1,BarthesBiesel:2016.1}.

Furthermore we use a bending-potential \cite{KruegerT:2016}
\begin{align}
	V_h = -\frac \kappa 2 \sum \limits_{i,j} \left( 1 - \cos \beta_{i,j} \right),
\end{align}
where $\beta_{i,j}$ is the angle between two  normal vectors of neighboring triangles of beads,
and a potential to conserve the volume of the capsule \cite{KruegerT:2016}
\begin{align}
	V_V = -\frac{k_v}{V_0} (V(t)-V_0)^2.
\end{align}
$V(t)$ means the Volume at a given time t and $V_0=\frac{4}{3}\pi R^3$ the desired Volume of the capsule (with the Radius R of the not deformed spherical capsule). The force in case of the capsule is given by $\bs F_i = -\nabla_i \left[ V_{h} + V_V+V_{NH}\right]$

If not stated otherwise we use the following parameters. 
For the flow, $u_0=0.5$, $d=60$, $\eta=1.0$; the ring, $a=0.5$, $k=0.175$, $\kappa=6.0$,  $N=16$, $b=2.5$ (which gives a ring radius 
of $R=6.36$); the capsule: $G=0.1$, $\kappa=0.1$, $k_v=3.0$, $a=0.4$, $N=642$, $b=1.0$ (which gives 
a capsule's radius  $R=6.6$); vertical force  $\bs F_g=-0.01\hat{{\bf e}}_y$.

\section{Qualitative explanation of cross-stream migration in vertical flows}

\begin{figure}[htb]
%\vspace{-1mm} \quad
\begin{center}
\includegraphics[width=0.9\columnwidth]{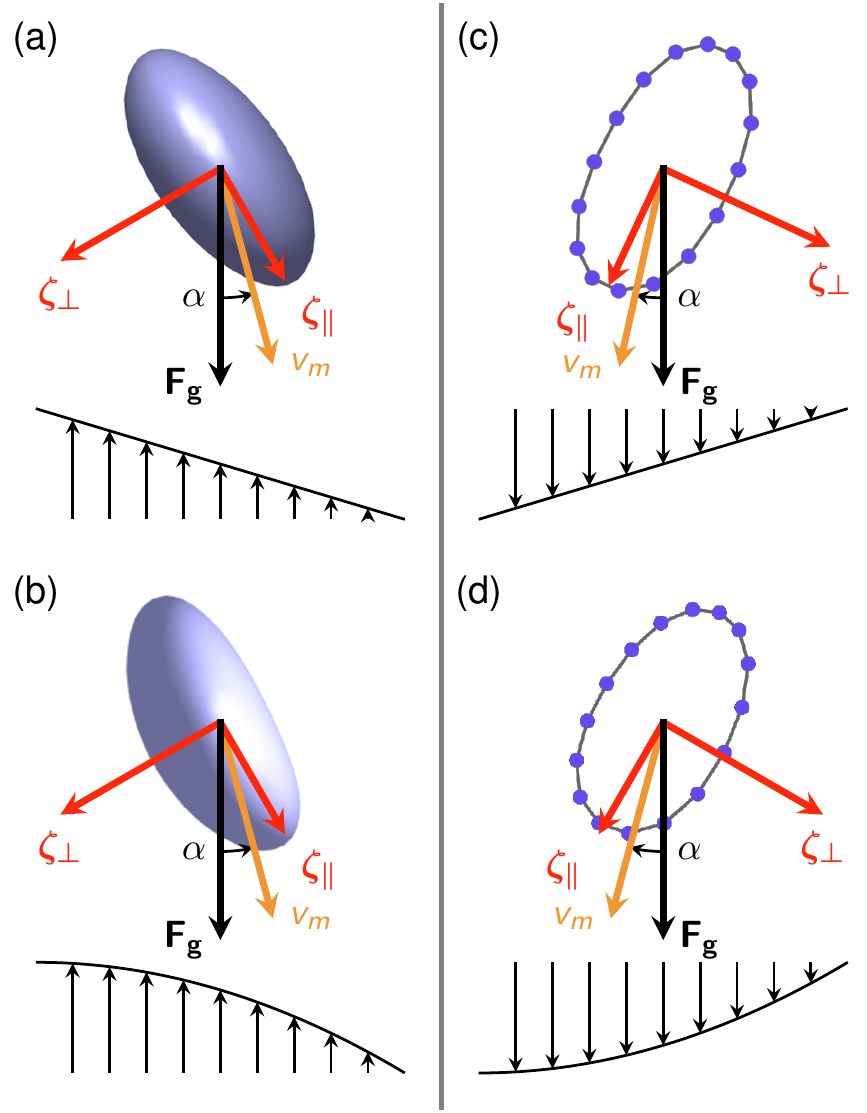}
\end{center}
\vspace{-4mm}
\caption{
Soft capsules and rings are deformed by shear flows as indicated in (a)-(d) for moderate local shear rates.
The particle's deformation in a linear shear and a Poiseuille flow 
is slightly different as indicated by the differences between the shapes in (a) and (b) for the capsule 
in (c) and (d) for the ring.  The particle's major axis in shear flow is inclined 
with respect to the flow lines and the particles has different Stokes drag coefficients $\zeta_\parallel$
parallel  and $\zeta_\perp$ perpendicular to the major axis. Therefore, 
 if an external force ${\bs F}_g$  acts on the particle (parallel or antiparallel to the flow direction) 
 the resulting particle migration velocity  $v_m$ encloses 
 an angle $\alpha$ with the straight flow lines and leads to cross-stream line migration (CSM).
 If  ${\bs F}_g$ points downward (upward) then the horizontal (cross-streamline) component of
 ${\bf v}_m$ and the shear gradient have opposite (equal) sign. Accordingly, for a given force ${\bs F}_g$ a reversal of the flow
 direction leads to a reversal of the shear gradient and therefore of  the particle's (horizontal) CSM direction.
} 
\label{Sketchmech}
\end{figure}

A capsule or a ring-polymer model  exposed to a linear shear or a Poiseuille flow is deformed by the local shear 
gradient, as shown in Fig. \ref{Sketchmech}, but
the  capsule's (ring's) shape is not identical 
in a linear shear and a Poiseuille flow (here demonstrated for moderate local shear rates).
However, for flows the capsule (ring) may be described in a first  approximation  
by a rotational symmetric ellipsoid (elliptical polymer) with a Stokes drag-coefficient 
$\zeta_\perp$ ($\zeta_\parallel$) in the direction  
perpendicular (parallel) to the major axis 
and  $\zeta_\perp > \zeta_\parallel$  \cite{BarthesBiesel:2016.1}.

The  major axis of a tank-treading capsule (ring) includes with the 
straight flow lines of both flows an angle, whereby the sign and the magnitude of this angle are determined by the sign and
the magnitude of the local shear rate \cite{BarthesBiesel:2016.1}.
The buoyancy force acting  on a particle points  upward for a light and downward for a heavy particle, i. e. 
it is either parallel or antiparallel to the flow lines. 
According to different drag coefficients, $\zeta_\perp> \zeta_\parallel$, 
an external force on an inclined capsule (ring) in shear flow causes 
an oblique migration velocity ${\bf v}_m$, as 
shown in Fig.~\ref{Sketchmech}. The inclination angle of the ellipsoid (ring) 
and therefore the inclination angle $\alpha$ 
of ${\bf v}_m$
depend on the sign of the local shear rate (see also analytical results below). 
For both flows a reversal of the flow direction leads to a reversal of
the local shear gradient and simultaneously to a reversal 
of the horizontal component ($x$ component) of the migration velocity ${\bf v}_m$: I. e. a  
reversal of the flow direction causes a reversal of the cross-streamline migration of non-neutrally buoyant soft particles.  
Furthermore,
if the buoyancy force is downward  (upward) then the shear gradient and the horizontal
migration direction are antiparallel (parallel).

 %\wznote{shear, wall}

\section{Small capsule-deformations}
%%%%%%%%%%%%%%%%%%%%%%%%%%%%%%%%%%%%%%%%%%%%%%%%%%%%%%%%
The shape of a Neo-Hookean capsule, its anisotropic drag and its cross-streamline 
drift in  a linear shear flow $\bs u_0=\dot \gamma x\hat{e}_y\ $ can 
be determined analytically in the range of a small 
capillary number $\mbox{Ca}=\dot \gamma \frac{\eta R}{G}$ \cite{BarthesBiesel:1981.1,BarthesBiesel:2016.1}.
In this limit, the  capsule shape is given by the equation 
\begin{eqnarray}
 r^2&=&x^2+y^2+z^2=R^2+\frac{5}{3} \frac{\mbox{Ca}}{\dot\gamma}\bs r^T \cdot \bs J \cdot \bs r+ O(\mbox{Ca}^2)\,,\nonumber \\
 \bs J&=&\frac{1}{2}\left[(\nabla\otimes\bs u_0)+(\nabla\otimes\bs u_0)^T\right]\,,\nonumber \\
 \bs r&=&(x,y,z)^T\,,
\end{eqnarray}
which describes  an ellipsoid with three different axes. The major axis forms with the undisturbed straight stream lines an angle of about $\frac{\pi}{4}$.
The length of the three axes are
\begin{align}
 d_{1,3}=&\frac{\sqrt{2}}{\sqrt{2\mp\frac{25}{3}\mbox{Ca}}}R\,, \quad \mbox{   (major/minor axis)} \nonumber \\
 d_2=&R\,. \label{eq_major_axis}
\end{align}
In order to proceed with analytical calculations we make a common approximation and assume rotational symmetry with 
respect to the major axis.  Then the length of the major/minor axis are given by
\begin{align}
 a=&\frac{d_2+d_3}{2}\,, \qquad \mbox{   (two minor axes)}\\
 b=&d_1\,. \qquad \qquad \mbox{   (major axis)}
\end{align}
The drag coefficients of a rotational symmetric ellipsoid parallel and perpendicular to the major axis 
are given by Perrin's formulas (see e.g. \cite{Brenner:1981} and references therein)
\begin{align}
\zeta_\perp=&\frac{8}{3}\frac{1}{\frac{\beta}{\beta^2-1}+\frac{(2\beta^2-3)\ln\left(\beta+\sqrt{\beta^2-1}\right)}{(\beta^2-1)^\frac{3}{2}}}\,, \label{eq_zeta_perp}\\
\zeta_\parallel=&\frac{8}{3}\frac{1}{\frac{2\beta}{1-\beta^2}+\frac{(2\beta^2-1)\ln\left(\frac{\beta+\sqrt{\beta^2-1}}{\beta-\sqrt{\beta^2-1}}\right)}{(\beta^2-1)^\frac{3}{2}}}\, \label{eq_zeta_para}
\end{align}
with $\beta=\frac{b}{a}$. The  migration velocity of the  ellipsoid perpendicular and parallel to the stream lines is obtained by decomposing
the buoyancy fore
$\bs F_g=F_g\hat {\bf e}_y$ into its component along the major axis, ${\bf F}_{g,\parallel}$, 
and perpendicular to it, ${\bf F}_{g,\perp}$.
 The migration velocity across the streamlines is then given by
\begin{equation}
 v_m=\left(\frac{\bs F_{g,\perp}}{\zeta_\perp}+\frac{\bs F_{g,\parallel}}{\zeta_\parallel}\right)\cdot \hat {\bf e}_x\,.
 \label{eq:vmf}
\end{equation}
A Taylor expansion with respect to Ca gives at leading order
\begin{equation}
 v_m=\frac{5}{96}\frac{F_g}{\pi\eta R}\mbox{Ca}+O(\mbox{Ca}^2)\ .
 \label{eq_vm_Ca}
\end{equation}
The capsule's shape in a Poiseuille flow at an 
off-center position has the shape of a slightly deformed ellipsoid,  as indicated also in Fig.~\ref{Sketchmech}. 
The deformation of a spherical shape is  determined by the local shear rate
in Poiseuille flow  at  the capsule's center $(x_c,y_c)$: $\dot \gamma= -\frac{2 u_0 x_c}{d^2}$.
With the local capillary number 
$\mbox{Ca}=\dot \gamma \frac{\eta R}{G} = -\frac{2 u_0 x_c \eta R}{d^2 G}$ one obtains within this approximation
the position dependent cross-streamline migration velocity in Poiseuille flow
\begin{equation}
 v_m\approx-\frac{5}{48}\frac{F_g u_0 x_c}{\pi G d^2}\,.
 \label{eq_vm_ana}
\end{equation}
The force induced  velocity $v_y=({\bf v}_m)_y$ relative to the unperturbed flow 
 can be calculated analogous as in a linear shear, which includes besides 
 result from the Stokes drag a deformation dependent correction proportional to $Ca$:
\begin{equation}
 v_y=\frac{F_g}{6\pi\eta R}+\frac{5}{288}\frac{F_g}{\pi\eta R}\mbox{Ca}+O(\mbox{Ca}^2)\, .
\end{equation}
Its explicit form for Poiseuille flow is then
\begin{equation}
 v_y\approx\frac{F_g}{6\pi\eta R}-\frac{5}{144}\frac{F_g u_0 x_c}{\pi d^2 G}\, .
\end{equation}
If the external force is antiparallel to the flow and if it is large enough,
 the capsule moves against the flow direction. This is approximately the case if $v_y$ is larger then the velocity of the Poiseuille flow at the capsule center
\begin{align}
v_y>&u_0(x_x,y_c)\\
 \frac{F_g}{6\pi\eta R}-\frac{5}{144}\frac{F_g u_0 x_c}{\pi d^2 G}>&u_0\left(1-\frac{x^2}{d^2}\right)\\ 
 F_g>&\frac{144\pi\eta R d^2 G u_0}{24 d^2 G-5R\eta u_0 x_c} \left(1-\frac{x_c^2}{d^2}\right)
\end{align}

%\wznote{Hier noch ein zwei Sätze, wieso wir das berechneten.}

\section{Numerical results on CSM in unbounded flows}

To verify our approximate analytical predictions and qualitative descriptions of the migration
of non-neutrally buoyant soft particles in unbounded flows, 
we simulate the Stokesian dynamics of capsules and a rings.

This ensures that the occurred migration is no inertial effect or imposed by wall interactions. Since the migration reversal does only depend on the local shear rate, we confine the numerical investigation, without loss of generality, to that of the Poiseuille flow.

The setup for all simulations can be seen in Fig. \ref{Sketchsink}. The imposed flow is a Poiseuille flow applied in y-direction with $-d<x<d$. The deformable particles are placed with center of mass positions $x_c$. As consequence, particles with negative migration velocity ($v_m<0$) will migration to the wall and positive migration means center migration.

\begin{figure}[htb]
\vspace{-2mm}
\begin{center}
\includegraphics[width=0.92\columnwidth]{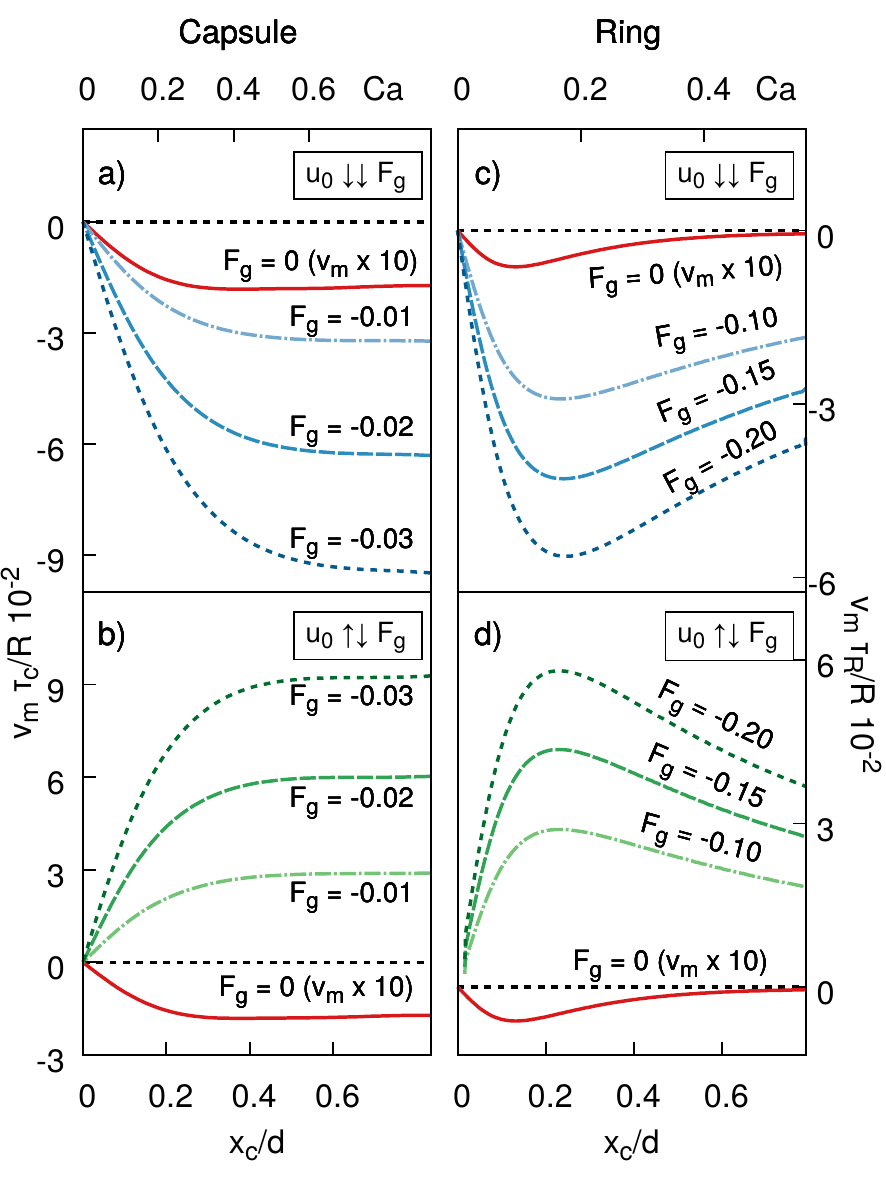}
\end{center}
\vspace{-6mm}
\caption{Cross-streamline migration-velocity $v_m$ of deformable capsules [a) and b)] and rings [c) and d)] in {\it unbounded}
vertical Poiseuille flows as function of the capillary number Ca and for 
different values of the vertical force $F_g$.  For parallel flow and force directions, 
${\bf u}_0 \downarrow \downarrow {\bf F}_g$, 
the particles migrate to the center of the Poiseuille flow, i. e. $v_m<0$ ($v_m>0)$ in range $x_c>0$ ($x_c<0$), only faster 
than in the case of neutral particles with ${\bf F}_g$.
For an antiparallel configuration, ${\bf u}_0\uparrow \downarrow {\bf F}_g$, the migration is reversed 
and  away from the center of Poiseuille flow. 
These results show, that force induced relative velocity ${\bf v}_m$ in Fig.~\ref{Sketchmech}
increases with the magnitude of $F_g$.
}
\label{v_m_Weissen_ring}
\end{figure}

As reference, we first discuss the case without external force, which is shown as red curve in Fig. \ref{v_m_Weissen_ring}. For this case we observe the already known center migration for both, the capsule and the ring [zitat: zentrumsmigration für soft-particle low Re]. 

If we now apply a gravitational force parallel to the stream lines ($\bs u_0 \downarrow \downarrow \bs F_g$), we see an increase of the center migration (Fig. \ref{v_m_Weissen_ring} (a) and (c)). This additional center-migration is enlarged for higher external forces. During the migration to the center, the migration speed decreases, which is a result of the position dependent shear-rate of the Poiseuille flow. When the particles reach the center line, $v_m$ vanishes and the particle follows the stream lines. 

For the case of antiparallel force and flow-direction ($\bs u_0 \uparrow \downarrow \bs F_g$), we find the opposite behavior (Fig. \ref{v_m_Weissen_ring} (b) and (d)). The migration direction is reversed and the particles experience a positive migration-velocity. This effect is also correlated to the force strength and position of the particle. The migration in this case does not vanish on its way remote from the center, since there is no repulsive interaction with the walls. 
This migration reversal is observed for both, the capsule and the ring. Both migration curves look similar, which means the reversal effect does not depend on the detail of the particle but is a more generic feature of the interplay between deformation and external forces. This is consistent with the results from \cite{Yeung_ES:2002.1}, where they could measure this behavior for bundled DNA-molecules, which have no impenetrable surface.

\section{Cross-stream drift  between walls}

\subsection{Migration of a sedimenting capsule}
\begin{figure}[!h]
\vspace{-2mm}
\begin{center}
\includegraphics[width=0.8\columnwidth]{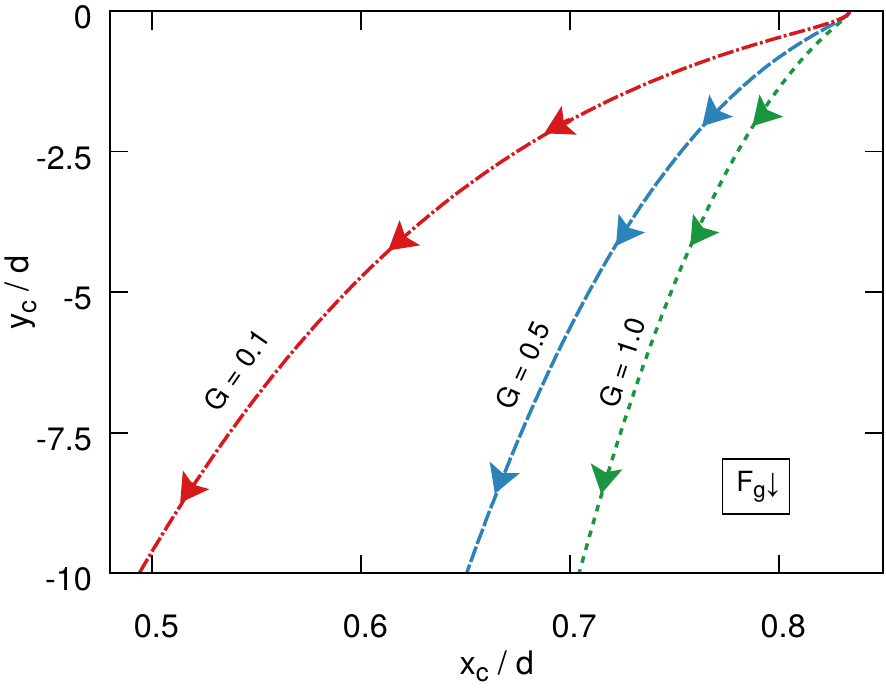}
\vspace{-4mm}
\end{center}
\caption{Trajectories of a deformable capsule sedimenting between two walls in the  absence of  flow ($\bs u_0=0$) 
for three different values of the Neo-Hookean stiffness $G$. 
Softer particles move faster away from the wall (at $\frac{x}{d}=1.0$) and
rigid particles don't.}
\label{Tra_sedimenting}
\end{figure}

The influence of the flat walls of the channel should also be studied. At first we investigate the interaction between a sedimenting capsule and the walls of the channel without a flow. We observe a repulsion of the capsule from the wall which depends on the stiffness of the capsule, see Fig. \ref{Tra_sedimenting}. The softer the capsule is the stronger is the repulsion. This is consistent with the fact that a solid particle sinks parallel to the wall (see zitat).

The reason of the repulsion is the deformability of the capsule. The part of the capsule closer to the wall is subjected to a friction with the wall (transmitted via the fluid) and lags behind. The other part of the capsule more away from the wall moves therefore faster. This stretches the capsule and lasts until a steady state is reached which has a shorter and a longer axis. The longer axis points away from the wall if seen from the center of the capsule in direction of the external force. This leads, as described above, to an asymmetric drag and the capsule moves not completely in direction of the force but a bit shifted towards the major axis. This means it drifts away from the wall. This mechanism is different from the wall repulsion of capsules in a shear flow without a external force where a lift force due to the tank treading leads to the wall repulsion.

\subsection{Dynamics of rings and capsules in Poiseuille flows}
\begin{figure}[htb]
\vspace{-2mm}
\begin{center}
 \includegraphics[width=0.92\columnwidth]{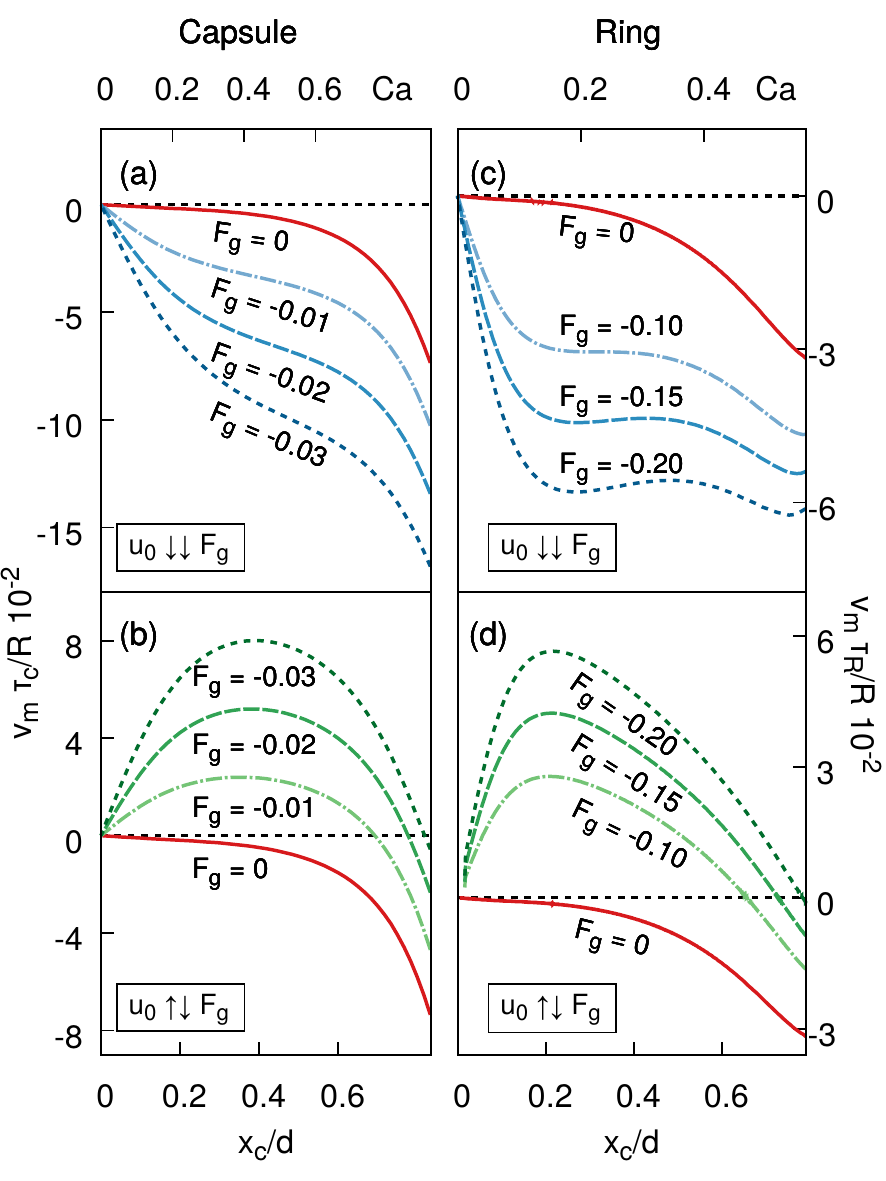}
\end{center}
\vspace{-6mm}
\caption{Migration velocity of soft particles analogous to Fig. 2.
with additional wall interaction. The migration towards the center ($v_m<0$), 
observed in the parallel case ($u_0 \downarrow \downarrow F$, upper picture) is increased near the wall. For the antiparallel case ($u_0\uparrow \downarrow F$, lower pictures), the repulsive wall interaction leads to stable positions ($v_m=0$) between the center and the wall. This effects are again stable for different types of particles and can be seen for both capsules and ring polymers.}
\label{v_m_Weissen_capsule}
\end{figure}

The contribution of the wall interaction to the migration of capsules in a Poiseuille flow is examined here. The Fig. \ref{v_m_Weissen_ring} and \ref{v_m_Weissen_capsule} show the migration velocity $v_m$ of a ring and a capsule as function of the capillary number Ca for different values of the force $F_g$ in case of an unbounded or an bounded flow. A comparison of both Fig. shows that far away from the walls the migration is similar in case with and without walls but is changed close to the walls where the repelling lift force becomes important. In case of a parallel external force and flow the lift force enhances the migration to the center. In case of an antiparallel external force and flow the bulk migration to the wall is hindered or surpassed by the lift force. The capsule and the ring migrates away from the wall if it is close to it. This leads to stable lateral position $x_{eq}$ outside the channel center where the migration due to the external force and due to the the lift force are equal.

\begin{figure}[htb]
\begin{center}
\includegraphics[width=0.8\columnwidth]{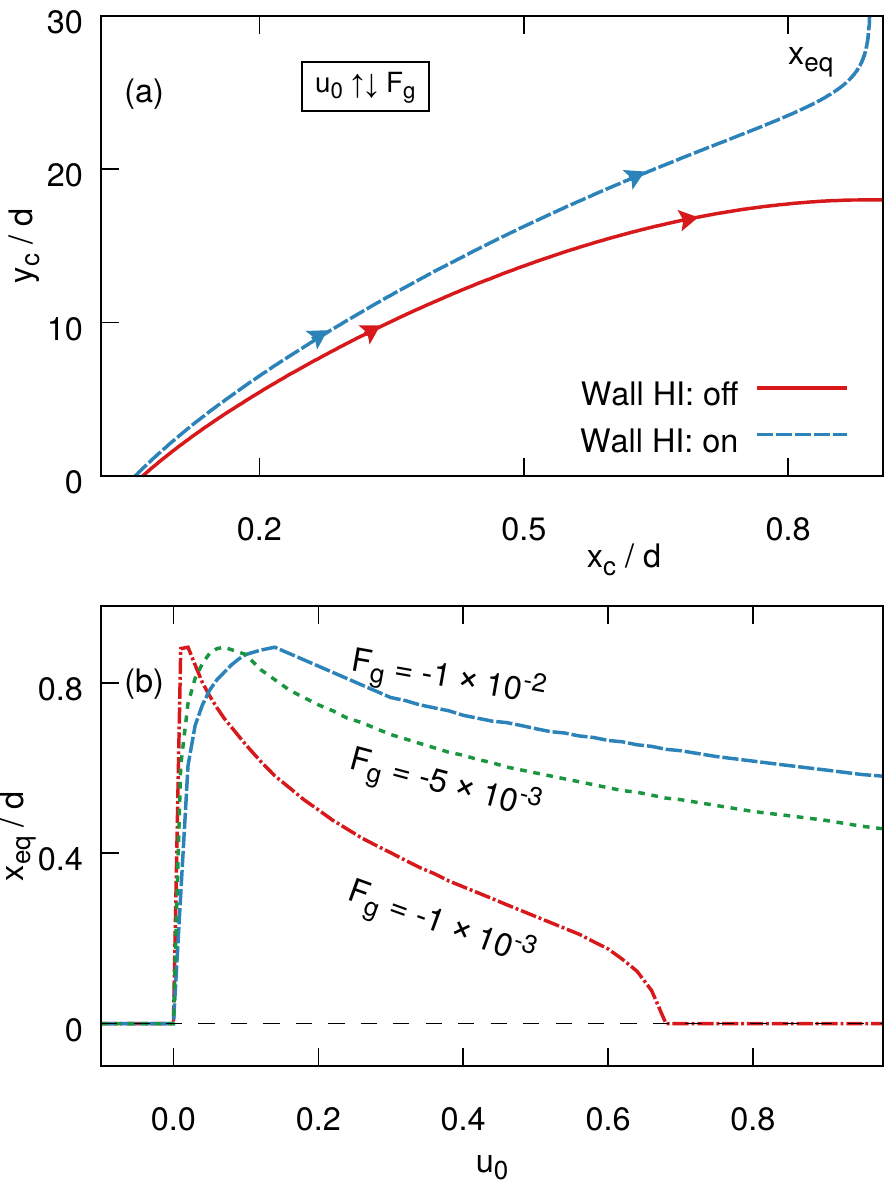}
\end{center}
\vspace{-6mm}
\caption{(Color online) (a) Trajectory of a deformable capsule immersed in a plane Poiseuille flow $\bs u_0(\bs r)$ between two parallel walls (see also \fig{Sketchsink}). The flow is antiparallel to the negative buoyancy force $\bs F_g$. Without the wall interaction (red) the capsule migrates towards the walls due to $\bs F_g$ until it collides with the wall. With wall interaction (green) the migration stops close to the walls at $\frac{x_c}{d}\approx0.85$ because the wall is repulsive (see \fig{Tra_sedimenting}). This means the capsule reaches a stable equilibrium distance to the wall $x_{eq}$.
(b) Equilibrium position $x_{eq}$ for a capsule as function of the flow velocity for three different values of the negative buoyancy forces. 
%Due to the symmetry $-x_{eq}$ is also an equilibrium position.
Negative values of $u_0$ correspond to a flow parallel to the force thus the center of the flow is the stable position. Positive values of $u_0$ mean flow and negative buoyancy force are antiparallel which allows stable off center positions. This leads to a transition of $x_{eq}$ from zero to non-zero values at $u_0=0$. Beyond a maximum at approximately $u_0\approx 0.05$ the stable position $x_{eq}$ is the closer to the center the faster the flow is because of the tank-treading motion. The tank-treading motion which leads to a center migration without $\bs F_g$ becomes stronger with higher $u_0$. At the values $F_g=-10^{-3}$ and $u_0>0.5$ the tank-treading dominates and the capsule migrates to the center, as in the case without $\bs F_g$.
Furthermore a higher force $\bs F_g$ means a stable position closer to wall
%The three curves display a maximum at around $u_0\approx 0.05$. The curve with $\bs F_g=-1\times 10^{-3}$ is zero at approximately $u_0>0.5$. The Migration to the center due to the tank-treading motion, which also exists without $\bs F_g$ is here stronger than the migration to the walls due to $\bs F_g$. Furthermore a higher force $\bs F_g$ means a stable position closer to wall.
}
\label{xequilibrium2}
\end{figure}

The drift towards such a stable wall distance at $x_{eq}\approx 0.85 d$ and the dependence of $x_{eq}$ on $u_0$ and $F_g$ is shown in Fig. \ref{xequilibrium2} . If the flow is parallel to the force $u_0<0$ the stable position is the channel center because the external force, the center migration occurring also without an external force and the wall repulsion lead to a migration to the center. If the flow is reversed (meaning now antiparallel to the force) the migration due to the force leads to a wall migration and stable, lateral off-center positions occur. They are at approximately $u_0=0.05$ the closest to the wall depending on the external force. Beyond this maximal off-center position $x_{eq}$ becomes closer to the center the higher $u_0$ is because the increasing tank-treading motion and $x_{eq}$ is the closer to the wall the larger the external force is. With a weak force and a strong flow e.g. $F_g=-10^{-3}$ and $u_0>0.7$ the stable position is again the center of the channel despite they are antiparallel.

\subsection{Comparison of simulation methods and analytical approximation}
We used in our simulations with the Oseen and Blake tensor the assumption that the external force acts on the surface of the capsule. In our analytical approximation we approximate the shape of the capsule as an rotational ellipsoid to calculate the drag. To justify these approximations we compare the results with an Lattice Boltzmann method (LBM) with the BGK collision operator and a singe relaxation time. Flow and particle are coupled by the immersed boundary method. With lattice Boltzmann, it is possible to simulate an external force that acts on the interior of the capsule or on the surface.

With the LBM we can compare the analytical approximation, the Oseen and Blake tensor simulation, a LBM simulation with an external force acting on the surface and a second LBM simulation with an external force acting on the interior of the capsule. We determine the migration velocity of a capsule with an external force in an antiparallel directed, bounded Poiseuille flow as function of the lateral position, see Fig. \ref{yc_vm_LBM.eps}. All simulations show that the capsule migrates towards the walls and stops at a certain distance to the wall due to the wall repulsion. Furthermore they all display zero migration in the center, a maximum of the migration in the middle between center and wall and a stable, stationary position at approximately $x_c\approx 0.85d$. So the simulations agree qualitatively. Quantitatively the simulations differ at the maximum of the migration velocity (approximately a factor of two) but agree well at low capillary numbers close to the center of the channel. This means the Oseen and Blake tensor describes the capsule qualitatively correct.

The analytical approximation agrees well at low capillary numbers at the center of the channel. This is due to the fact that small deformations are assumed. Also the analytical approximation can not reproduce the stationary position because the wall repulsion is not included here. This means the analytical approximations is justified as long as the assumptions of a small capillary number and a position far away from the wall are given.
\begin{figure}[htb]
\begin{center}
\includegraphics[width=0.75\columnwidth]{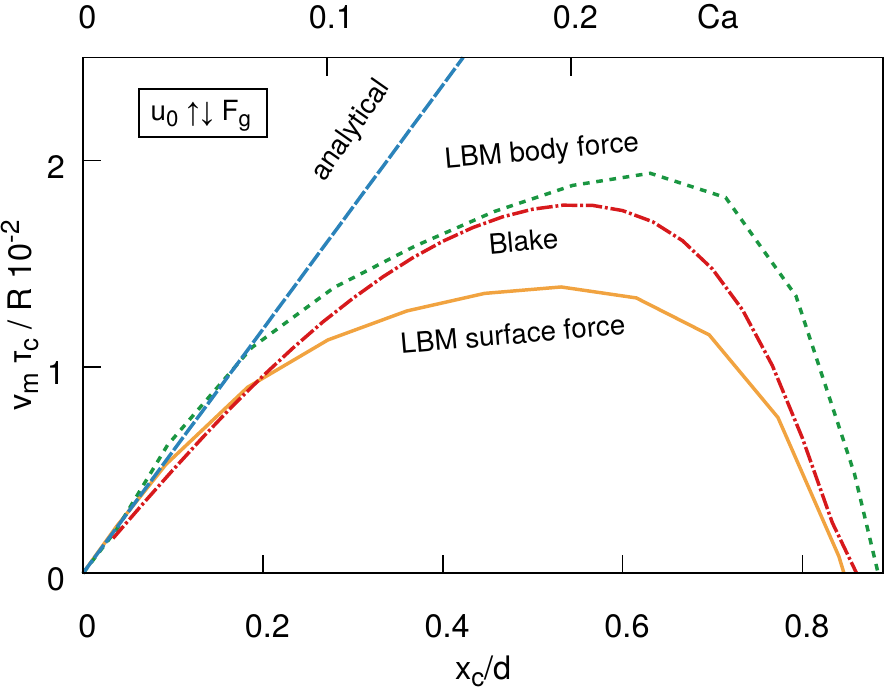}
\end{center}
\vspace{-4mm}
\caption{The migration velocity $v_m$ of a capsule as function of the lateral position $x_c$. We compare four different methods: Simulations with the Oseen and Blake-Tensor (purple) where $\bs F_g$ acts on the beads (surface of capsule), Lattice-Boltzmann simulations with $\bs F_g$ acts on the surface of capsule (red) or on the interior volume (green) and an analytical approximation (blue, see eq. \ref{eq_vm_ana}). All the simulations show the same qualitative behavior. The approximation of the Oseen-Tensor and the force $\bs F_g$ acting on the beads does not change the qualitative behavior of the capsule, just the quantitative values. Also the calculation fits the simulations as long as the assumption of small Ca and a position far away from the wall (negligible wall interaction) is valid. We used the parameters $d=60$, 
$u0=0.01$, 
$\tau=1.0$, 
Density of the Fluid $\rho=1.0$, 
$\eta=1.0/6.0$, 
Channel size in flow direction $N_x=100$, 
Channel size in z-direction $N_z=100$, 
$\kappa=0.001$, 
$G=0.001$, 
$k_v=0.01$, 
$b=1$, 
$\bs F_g=-10^{-4}\hat{e}_x$,
$a=0.2$.
}
\label{comparison}
\end{figure}

%Parameters LBM used for \fig{comparison}:
%$d=60$, 
%$u0=0.01$, 
%$\tau=1.0$, 
%Density of the Fluid $\rho=1.0$, 
%$\eta=1.0/6.0$, 
%Channel size in flow direction $N_x=100$, 
%Channel size in z-direction $N_z=100$, 
%$\kappa=0.001$, 
%$G=0.001$, 
%$k_v=0.01$, 
%$b=1$, 
%$\bs F_g=-10^{-4}\hat{e}_x$.
%Parameters Oseen: same as LBM-Parameters with bead size $a=0.2$.
%   d=60
%   u0=0.01
%   eta=1.0/6.0	
%   k_v=0.01
%   kappa=0.001
%   b=1.0
%   Gs=0.001
%   double F_ext_x=-0.0001

\section{Summary and conclusions}

The migration of the sinking (elevating) soft particles away from the walls has a similar origin as the lift force observed
in linear shear flow \cite{Misbah:1999.1,Seifert:1999.1,Viallat:2002.1}.\\

Migration during sedimentation\\

This cross-streamline migration is more efficient than bulk migration.\\

By this method we can separate particles with respect to their density and stiffness.\\

Shear flow\\

\acknowledgments

\end{document}